\documentclass[aps,
prd,
twocolumn,showpacs,nofootinbib,10pt]{revtex4-2}
\usepackage[table]{xcolor}
\usepackage{graphicx} \usepackage{amsmath} \usepackage{amssymb}
\usepackage{amsfonts} \usepackage{bm}
\usepackage{array}
\usepackage{siunitx}

\usepackage[justification   = raggedright,singlelinecheck=false]{caption}

\usepackage{ytableau}
\usepackage{multirow}
\usepackage{mathtools}
\usepackage{tikz}
\usepackage{fancyhdr}
\usepackage{tabularray}
	
\fancypagestyle{specialfooter}{%
\fancyhf{}

\fancyfoot[R]{ \noindent\fbox{%
\parbox{\textwidth}{%
{\footnotesize \bf \tiny{Electronic version of an article published as   International Journal of Quantum Information Vol. 22, No. 07, 2450032 (2024)[10.1142/S0219749924500321] © [copyright World Scientific Publishing Company] [https://www.worldscientific.com/worldscinet/ijqi]}}
}
}}
}

\begin{document}

\newcommand{\be}{\begin{equation}} \newcommand{\ee}{\end{equation}}
\newcommand{\bea}{\begin{eqnarray}}\newcommand{\eea}{\end{eqnarray}}

\title{Quantum walk search for  exceptional configurations}

\author{Pulak Ranjan Giri} \email{pu-giri@kddi-research.jp}

\affiliation{KDDI Research,  Inc.,  Fujimino-shi, Saitama, Japan}

\begin{abstract} 
There exist  two types   of  configurations of marked vertices on a two-dimensional grid, known as  the  {\it exceptional configurations}, which are hard to find by the   discrete-time    quantum walk  algorithms.   In this article, we provide a  comparative  study  of  the quantum walk algorithm  with different coins  to search  these  {\it exceptional configurations} on   a two-dimensional grid.      We further  extend the analysis to   the hypercube, where only one type of  {\it exceptional configurations} are  present.  Our observation,  backed by numerical results, is  that    our  recently proposed modified coin operator  is the only coin  which can search both types of {\it exceptional configurations} as well as non-{\it exceptional configurations}   successfully. As a consequence, we observe  that  the existence of {\it exceptional configurations} are  not a quantum phenomenon, rather a mere  limitation  of some  of the coin operators. 

 \end{abstract}


\date{\today}

\maketitle\thispagestyle{specialfooter}

\maketitle 



\section{Introduction} \label{in}

Searching a database  is one of the important tasks in many applications of  computer science.  Given an unsorted database of size $N$, a classical computer  takes   $\mathcal{O}(N)$  time to find out a specific  element from the database  in the worst case scenario.    However, quantum computer can do the same job faster \cite{ni} than the classical computer, because it can exploit quantum superposition in its favour.   Grover's quantum search algorithm \cite{grover1} can search for a marked  element in the database quadratically faster, in  $\mathcal{O}(\sqrt{N})$  time.  A detailed review on Grover's   search algorithm and its generalizations  is presented in ref.  \cite{giri}.

Generalization of the Grover's  search to the  quantum search on  graph   is not straightforward.   Searching on graphs is subject to the constraint that  we are only allowed to shift  from one vertex of the graph to only the  nearest  neighbour vertices  in one  time step.  Although Grover search requires  $\mathcal{O}(\sqrt{N})$  iterations, but each  iteration needs another  $\mathcal{O}(\sqrt{N})$ time to perform all the reflections, making the total time  $\mathcal{O}(N)$ \cite{beni}. 

In this regard, quantum walk \cite{portugal}, which is the quantum counterpart of the classical random walk,  has been very much successful in searching on  graphs   faster than the classical exhaustive search.   Both  continuous-  \cite{childs} and discrete-time \cite{amba2} quantum walk can be exploited to search for  the  marked vertices on  a  graph.  Some of the examples of  the  quantum walk search include searching  on   one- \cite{lovet,giriijqi}, two- \cite{girimpa}  and  more than two-dimensional grid  \cite{childs1}  with periodic boundary conditions and many other graphs \cite{amba4,meyer} as well.  Specifically,   on two-dimensional grid  Grover's algorithm together with multi-level recursion \cite{amba1} can  search for  a marked vertex in  $\mathcal{O}(\sqrt{N}\log^2 N)$ time.  The discrete-time quantum walk can   perform the same  search in  $\mathcal{O}(\sqrt{N}\log N)$ time  with a possible improvement of  $\mathcal{O}(\sqrt{\log N})$ \cite{tulsi}  by some additional techniques.  However, lackadaisical quantum walk \cite{wong1,wong2,wong3} can directly  search on the two-dimensional grid in  $\mathcal{O}(\sqrt{N\log N})$ time without the help of any additional technique.    Optimal speed of  $\mathcal{O} (\sqrt{\frac{N}{M}})$  can be achieved by lackadaisical quantum walk search \cite{giri3} if additional long range edges of a specific type  are attached with the two-dimensional grid.   In continuous-time quantum walk,  optimal speed of $\mathcal{O} (\sqrt{N})$  for a single vertex search has  been achieved \cite{tomo}  by adding long range edges to the  two-dimensional grid.

In   discrete-time quantum walk,    marked  vertices are distinguished   from the  unmarked vertices by modifying the coin operator.  SKW coin,  $\mathcal{C}_{SKW}$, \cite{she} and Grover coin, $\mathcal{C}_{Grov}$, \cite{amba1}  are two such  widely used  modified coins, which   differentiate  marked vertices from the unmarked vertices.   For example, SKW  coin  based quantum walk search algorithm  applies Grover diffusion operator,  $D_0$,  to the  unmarked vertices and $-\mathbb{I}$ to the marked vertices.  Whereas,  Grover coin based quantum walk  algorithm applies    $D_0$ to the unmarked vertices and   $-D_0$ to the marked vertices.  Search algorithms based on these two  coins   have been  very  much successful in finding   a single  marked vertex  on  a graph.   
However, while searching for multiple marked vertices \cite{rivosh,nahi,saha} there are certain limitations to  the search problem.   For example, Grover coin  can not find two adjacent  marked vertices \cite{rivosh1}  on a two-dimensional periodic lattice.   Also,  the set of marked vertices,   arranged in a   $2k \times m$ or  $k \times 2m$ block  for any positive $k$ and  $m$,   can not  be found.     
However,   for  the set of  marked vertices  arranged in a  $k \times m$ block  for   $k$ and $m$  both  being odd,  Grover coin  can be used to  search any  of the marked vertices in  these blocks. 
On the other hand,  marked vertices,  arranged along the diagonal of a  two-dimensional grid  and its certain   generalised configurations \cite{amba5,men},    can not be searched using SKW coin.   Such configurations of marked vertices on the two-dimensional grid as well as on some other graphs, which can not be found by  the SKW and/or  Grover coin,   are collectively known as the {\it exceptional configurations}.

To overcome the limitations of the SKW and Grover coin, recently a  coin  operator,  $\mathcal{C}_{G}$,   is proposed by us  in ref. \cite{giriepjd}, which  is obtained by  modifying  the  coin operator,  $\mathcal{C}_{l}$,  of the lackadaisical quantum walk search.   Instead of flipping the sign of all the  basis states of the  marked vertices it only flips  the sign of the self-loop attached to the marked vertices followed by   the Grover diffusion operator acting on  the coin  basis states  of the  marked vertices.   Quantum walk search then effectively becomes  a  Grover search for the self-loops of the marked  vertices.  This coin can search the {\it exceptional configurations} as well as  any other type of configurations of the marked vertices.  

\begin{figure}[h!]
  \centering
     \includegraphics[width=0.40\textwidth]{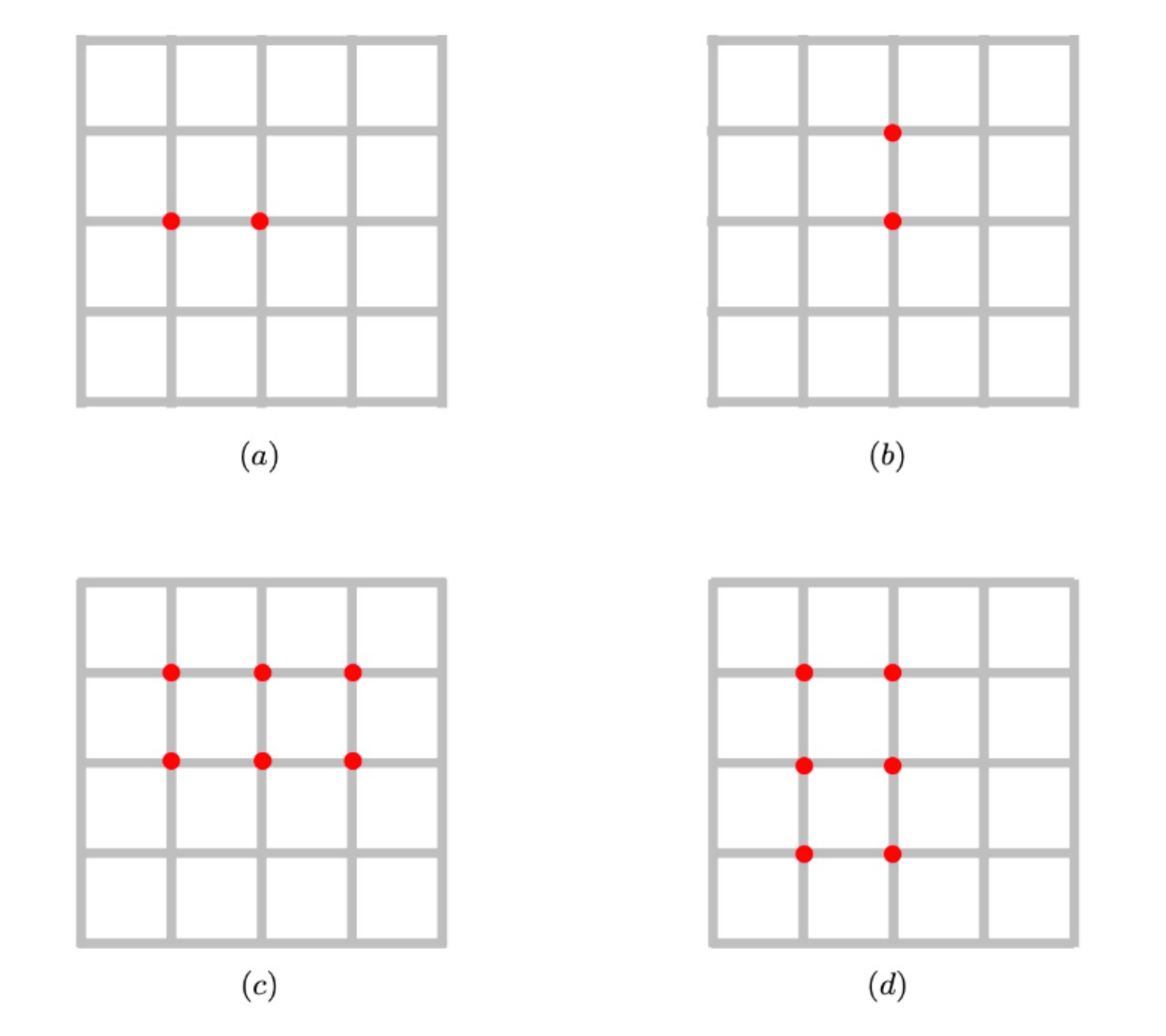}
          
       \caption{ Two-dimensional grid  with periodic boundary conditions:  a pair of marked  vertices in (a) horizontal direction, and  in  (b) vertical  direction; and a cluster  of marked  vertices of the form  (c) $k \times 2m $ and  (d) $2k \times m$, for $k,m$ being positive integers.}
\end{figure}

It has been seen that,   mostly SKW, Grover coin  and  coin for the  lackadaisical quantum walk search, $\mathcal{C}_l$,   have difficulty in   finding the  {\it exceptional configurations} of  marked vertices.  However,   adding more marked vertices makes  the classical  search problem  easer by making the search  time less and success probability more.  Based on this observation,  it was  suggested  in refs.  \cite{rivosh1,wong4} that    the  occurrence of the  {\it exceptional configurations} are  a  quantum phenomenon.

In this article a comparative study on the performance of quantum walk  algorithm  with different coin operators to search  {\it exceptional configurations} shows that,  the occurrence of {\it exceptional configurations} are merely  the  limitation  of some of the  coin operators, which can be overcome by a suitably modified coin operator and therefore not a quantum phenomenon. 
We  show that  the stationary state \cite{wong4}, which is responsible for the failure to search    {\it exceptional configurations} of a specific type by the Grover coin, actually does not remain stationary  under   $\mathcal{C}_{SKW}$ and $\mathcal{C}_{G}$ coins.    For the diagonal and its generalised form of {\it exceptional configurations},   we show that $\mathcal{C}_{l}$ and   $\mathcal{C}_{G}$  can successfully search the marked vertices in these configurations.  We have also  studied the quantum walk  search for  {\it exceptional configurations} on a hypercube.  Experimental results presented in this article  show that   $\mathcal{C}_{G}$  outperforms  rest of   coin operators discussed  in this article.

We arrange  this article  in the following fashion:  A brief discussion on the  quantum walk  search   on a two-dimensional periodic lattice is  presented  in section    \ref{qw}.   
Then {\it exceptional configurations}  and their quantum walk  search algorithm are   discussed in  section \ref{ex}.  Quantum walk search for the {\it exceptional configurations} on  the hypercube  is discussed in  section \ref{hyp}   and finally  we conclude in section \ref{con}.

\section{ Quantum walk  search on two-dimensional grid} \label{qw}
In this section,  we provide a brief discussion on  the  discrete-time quantum walk search algorithm  on a two-dimensional grid of size  $\sqrt{N} \times \sqrt{N}$ with periodic boundary conditions.  Each vertex has four standard  edges  represented  by the basis states $|c_x^{+}\rangle, |c_x^{-}\rangle,  |c_y^{+}\rangle, |c_y^{-}\rangle$ of the coin space.   Hilbert space of the graph,  $\mathcal{H}_G = \mathcal{H}_C \otimes \mathcal{H}_V$,   is the tensor product of  the  coin,   $\mathcal{H}_C$,  and   vertex,   $\mathcal{H}_V$,  space  respectively.  Quantum walk search starts with the initial state, $|\psi_{in}\rangle = |\psi_{c}\rangle  \otimes |\psi_{v}\rangle$,  prepared with the uniform superposition of all the basis states on both    coin and vertex  spaces  respectively as 
\begin{eqnarray}
 |\psi_{c}\rangle   &=& \frac{1}{\sqrt{4}} \left(  |c_x^{+}\rangle + |c_x^{-}\rangle+ 
 |c_y^{+}\rangle + |c_y^{-}\rangle   \right),\\
 |\psi_{v}\rangle &=&   \frac{1}{\sqrt{N}} 
\sum^{\sqrt{N}-1}_{{v_x, v_y} = 0} |v_x,  v_y\rangle \,.
 \label{in}
\end{eqnarray}
Time evolution operator  $\mathcal{U}$   then acts on   the  initial state   $|\psi_{in}\rangle$   repeatedly until the  final  state  
\begin{eqnarray}
|\psi_{f}\rangle =    \mathcal{U} ^{t} |\psi_{in }\rangle\,,
\label{fin}
\end{eqnarray}
has  high fidelity with  the state representing  the marked vertices.
Usually,     in the lackadaisical quantum walk   the final state  $|\psi_{f}\rangle$  directly  reaches to the  state for the  marked vertex. However,  in quantum walk search without  self-loop,  final state does  not have  high overlap with the state of  marked vertex. So,  we need amplitude amplification technique to achieve high and constant success probability.

The time evolution operator  in the quantum walk search algorithm  is  composed of the coin operator  $\mathcal{C}$   followed by the   flip-flop shift operator  $S$  as 
\begin{eqnarray}
 \mathcal{U} = S \mathcal{C}\,.
\label{uqw}
\end{eqnarray}
Depending on   how the coin operator acts on the  marked  and unmarked  vertices, it  can be given a general form 
\begin{eqnarray}
\mathcal{C}=  C_+ \otimes \left( \mathbb{I}_{N \times N} -    \sum_{i=1}^M |t_i \rangle \langle t_i |  \right) + C_- \otimes  \sum_{i=1}^M |t_i \rangle \langle t_i | \,, 
\label{qcgen}
\end{eqnarray}
where $C_+$  and $C_-$ act on the unmarked and  marked vertices respectively and $t_is$ are the marked vertices.   Two of the well known   coin operators obtained from eq. (\ref{qcgen})  are the following
\begin{eqnarray}
\mathcal{C}_{Grov} =  D_0 \otimes \left( \mathbb{I}_{N \times N} -   2 \sum_{i=1}^M |t_i \rangle \langle t_i |  \right) \,,  \hspace{1.8cm}
\label{qcgen1}\\  \nonumber
\mathcal{C}_{SKW} =   \hspace{6.5cm} \\  
D_0 \otimes \left( \mathbb{I}_{N \times N} -    \sum_{i=1}^M |t_i \rangle \langle t_i |  \right) - \mathbb{I}_{4 \times 4} \otimes  \sum_{i=1}^M |t_i \rangle \langle t_i | \,, 
\label{qcgen2}
\end{eqnarray}
where  the Grover coin,  $\mathcal{C}_{Grov}$,  is   be obtained by replacing $C_+ =   -C_-  =  D_0 = 2|\psi_{c}\rangle \langle \psi_{c}| - \mathbb{I}_{4\times 4}$  and  the SKW coin,  $\mathcal{C}_{SKW}$,  is obtained by replacing  $C_+ = D_0$,  $C_- = - \mathbb{I}$   in eq. (\ref{qcgen}).    Shenvi et al. \cite{she}  first used   $\mathcal{C}_{SKW}$ coin to  search for a marked vertex   on a hypercube in optimum time.   Both of the coins in eq.  (\ref{qcgen1}) and 
(\ref{qcgen2}) have been used  in  ref. \cite{amba2} to study spatial   search on a two  and higher dimensional grid, hypercube  and  on a complete graph.   Note that   these two coins    have been referred to as  Grover and  AKR coins of the AKR algorithm  respectively  in  ref. \cite{rivosh1}.   Previously, we  referred Grover coin as AKR coin in \cite{giriepjd}, because it has been used by  AKR  \cite{amba2}  to search for a marked vertex  on a complete graph.

Quantum walk  can be  generalised to lackadaisical quantum walk by  adding  a self-loop   state  $|c_l \rangle$ with weight $l$ to each vertex of the graph. Then the  coin operator for the lackadaisical quantum walk search  is given by 
\begin{eqnarray}
\mathcal{C}_l  =  D_l \otimes \left( \mathbb{I}_{N \times N} -   2 \sum_{i=1}^M |t_i \rangle \langle t_i |  \right)\,,
\label{qcgen3}
\end{eqnarray}
where the Grover diffusion operator  $D_l = 2|\psi_{c}^l\rangle \langle \psi_{c}^l | - \mathbb{I}_{5 \times 5}$  can be obtained from the initial state 
\begin{equation}
 |\psi_{c}^l \rangle   = \frac{1}{\sqrt{4+ l}} \left( |c_x^{+}\rangle + |c_x^{-}\rangle+ 
 |c_y^{+}\rangle + |c_y^{-}\rangle   + \sqrt{l}| c_l \rangle \right) \,.
 \label{in4}
\end{equation}
Note that  regular quantum walk  can be obtained from lackadaisical quantum walk  by setting   $l = 0$.  Usually, we need to find out an optimal value for the self-loop weight  $l$ for which the success probability is maximised. However, in this article we will only choose a  suitable  value  for the self-loop weight such that we have a reasonably hight success probability to carry out our analysis on {\it exceptional configurations}.

To overcome the limitations of the above mentioned coin operators in  searching  for the {\it exceptional  configurations}  of marked vertices,   we recently  proposed \cite{giriepjd} the following modified coin  
\begin{eqnarray} \nonumber
\mathcal{C}_{G}=   \left(D_l \otimes \mathbb{I}_{N \times N} \right)  \hspace{4.5cm} \\  
  \left( \mathbb{I}_{5 \times 5}  \otimes \mathbb{I}_{N \times N}-   2 \sum_{i=1}^M | c_l, t_i \rangle \langle  c_l, t_i |  \right)\,.
 \label{qc3}
\end{eqnarray}
After the coin operation,  the flip-flop shift  operator acts on the basis states of the vertex space as  
\begin{eqnarray}
S| c_x^{+} \rangle| v_x; v_y \rangle &=& | c_x^{-} \rangle| v_x + 1; v_y\rangle\,, \\
S| c_x^{-} \rangle| v_x; v_y \rangle &=& | c_x^{+} \rangle| v_x - 1; v_y\rangle\,, \\
S| c_y^{+} \rangle| v_x; v_y \rangle &=& | c_y^{-} \rangle| v_x; v_y + 1\rangle\,, \\
S| c_y^{-} \rangle| v_x; v_y \rangle &=& | c_y^{+} \rangle| v_x; v_y - 1\rangle\,, \\
S| c_l \rangle| v_x; v_y \rangle &=& | c_l \rangle| v_x; v_y \rangle\,.
\label{2Dcgrov}
\end{eqnarray}
The success probability to find the marked vertices  becomes 
\begin{eqnarray} 
p_{succ} = \sum_{i=1}^{M} |\langle t_i | U^t |\psi_{in}^l \rangle|^2\,,
 \label{psucc}
 \end{eqnarray}
where sum over coin basis states is also done  at the marked vertices. 

\begin{figure}[h!]
  \centering
     \includegraphics[width=0.45\textwidth]{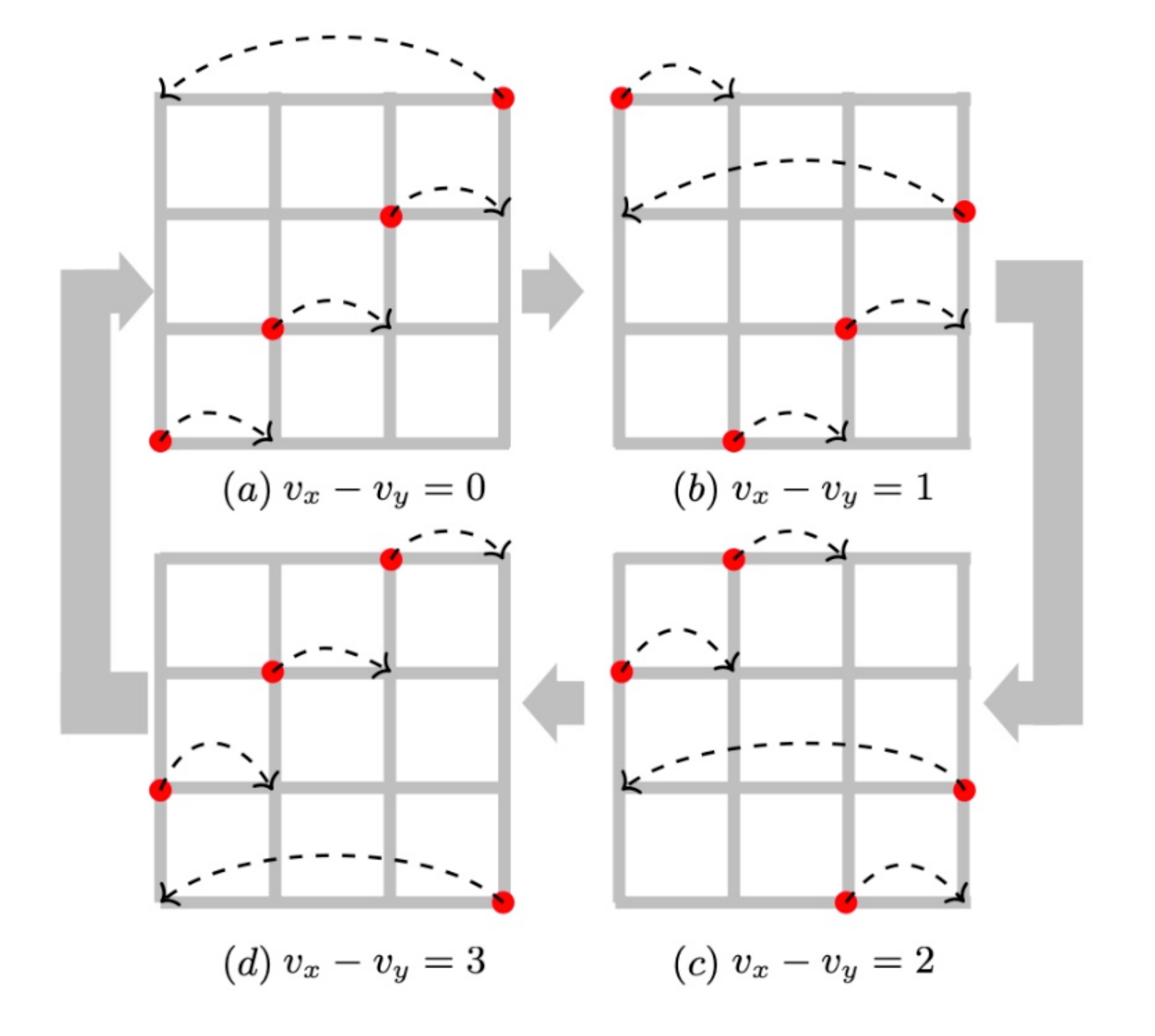}
          
       \caption{ Two-dimensional grid  with periodic boundary conditions with {\it exceptional configurations} of the  form  $(a) -(d)$  $i-j = const$.}
\end{figure}

\section{Exceptional configurations} \label{ex}
As stated in the introduction,  there are  {\it exceptional configurations} of marked vertices in the two-dimensional grid and in other graphs, for which the success probability to measure the marked vertices  does not improve despite repeated application of the iteration operator of the quantum walk search algorithm. Two types of {\it exceptional  configurations}, let us denote them as   $I$ and $II$,  are  reported in the literature \cite{rivosh1,amba5,men}.  In this section we will discuss these   configurations in detail and show how we  can search these  configurations by a suitably modified   coin operator. We also discuss the  existence of stationary states with  very hight  overlap with the initial  state of uniform superposition of all the basis states, which is responsible for the   failure to find    {\it exceptional   configurations I} by the  Grover coin based quantum walk search algorithm.

\subsection{Exceptional configurations I} \label{ex1}
These type of {\it exceptional configurations}  occur when two marked vertices are adjacent to each other.  Fig. 1  (a) and (b) are examples of   two  pairs  of adjacent marked vertices located in horizontal and vertical directions respectively.  More generally,    group of marked vertices of the form  $k \times 2m $ and   $2k \times m$, for $k,m$ being positive integers, are  {\it exceptional configurations} on a two-dimensional grid, which the  Grover coin can not find.  Fig 1(c) and (d),  with  marked vertices of the form  $3\times 2$ and $2\times 3$  respectively, are examples   of such general configurations.  

\begin{figure}[h!]
  \centering
     \includegraphics[width=0.40\textwidth]{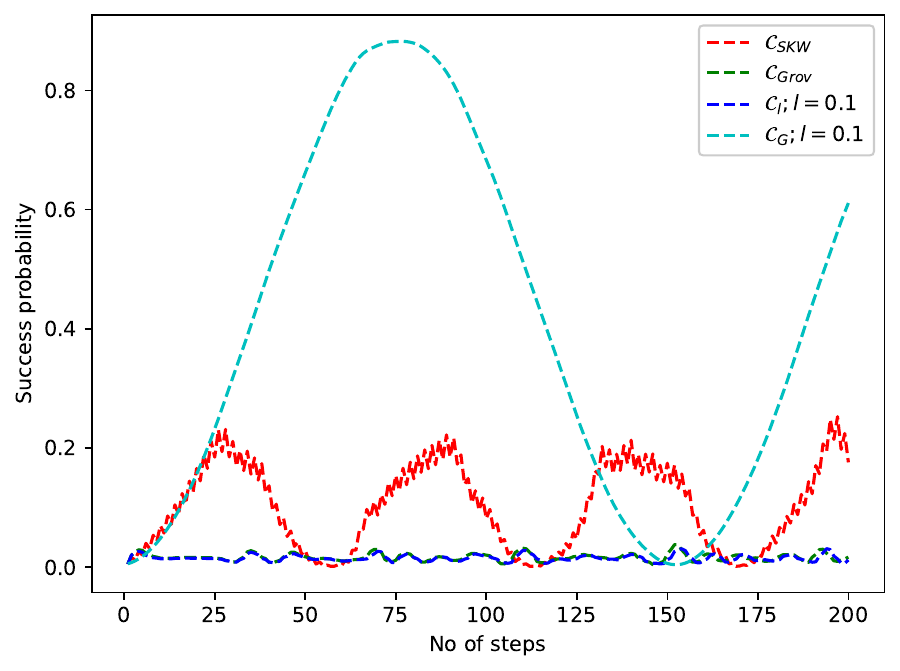}
          
       \caption{Success probability to measure a randomly chosen pair of  adjacent marked vertices in horizontal direction({\it exceptional configurations I})  on  a $20 \times 20$ square grid as a function of the number of iteration steps  for  four different coin operators.}

\end{figure}

\begin{figure}[h!]
  \centering
     \includegraphics[width=0.40\textwidth]{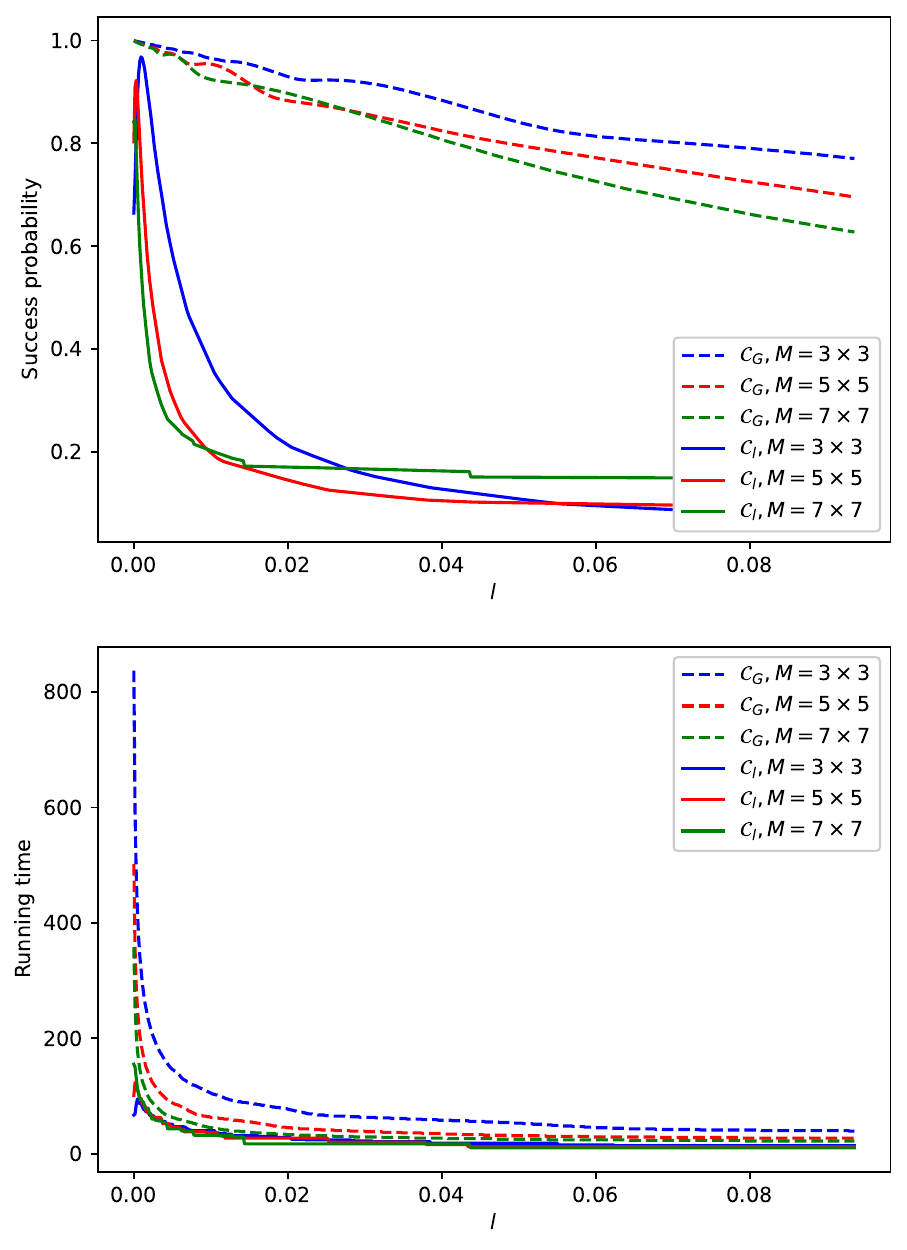}
          
       \caption{Success probability(top) and running time(bottom)  to search marked vertices of the block form $k\times m$, $k,m$ being odd,   as a function of the self-loop weight $l$   for a  $20 \times 20$ square grid.}

\end{figure}

\subsubsection{Stationary state} \label{ex2}
We are interested in the stationary state,  which has large overlap with the initial state of the quantum walk search.   For a detailed analysis on how to construct the stationary state and  use it in the quantum walk search to show that the  success probability  of the adjacent vertices   remains  close to their  initial success probability see refs. \cite{nahi,rivosh1,wong4}. 
To summarise the result for our purpose, let us consider a pair of  adjacent marked vertices on  a $\sqrt{N} \times \sqrt{N}$  square lattice.  These  two marked vertices can either be in horizontal direction as  $(i,j)$ and $(i+1,j)$   or in vertical direction as  $(i,j)$ and $(i,j+1)$,  where $i$ and $j$ are the coordinates representing horizontal and vertical directions respectively.  Since we are looking for a  stationary state, which is very close to the uniform superposition of basis states, we assume that all the vertices,  except the marked  ones, have the coin state of the form  in eq.  (\ref{in4}). 
However for the pair of marked vertices in the  horizontal direction we need  the following pair of  non-normalized coin states  
\begin{eqnarray} 
 |\psi_{c}^{x+}\rangle &=& |\psi_{c}^l \rangle   -\sqrt{4+l}  |c_x^{+}\rangle  \\ 
 |\psi_{c}^{x-}\rangle &=& |\psi_{c}^l \rangle   -\sqrt{4+l}  |c_x^{-}\rangle\,.
\label{inx}
\end{eqnarray}
Similarly,   for the pair of marked vertices in the  vertical  direction we need  the following pair of  non-normalized  coin states  
\begin{eqnarray} 
 |\psi_{c}^{y+}\rangle &=& |\psi_{c}^l \rangle   -\sqrt{4+l}  |c_y^{+}\rangle  \\ 
 |\psi_{c}^{y-}\rangle &=& |\psi_{c}^l \rangle   -\sqrt{4+l}  |c_y^{-}\rangle\,.
\label{iny}
\end{eqnarray}
Note that,  the above four states  are orthogonal to   $ |\psi_{c}^l\rangle $, i.e.,   $  \langle \psi_{c}^{x+}|\psi_{c}^l \rangle =  \langle \psi_{c}^{x-}|\psi_{c}^l \rangle =  \langle \psi_{c}^{y+}|\psi_{c}^l \rangle =  \langle \psi_{c}^{y-}|\psi_{c}^l\rangle =0 $.

{\it Case 1:} The stationary state in this case has  all the unmarked vertices  with coin state of the form  $ |\psi_{c}^l\rangle $ and  the two horizontal  marked vertices  $(i,j)$ and $(i+1,j)$  with their coin states   of the form   $|\psi_{c}^{x+}\rangle$ and 
$ |\psi_{c}^{x-}\rangle $ respectively.  Specifically,  the stationary state   has the following form 
\begin{eqnarray} \nonumber 
|\psi_{hstat}\rangle =   |\psi_{in}^l \rangle   \hspace{5cm}\\
 -\sqrt{\frac{4+l}{N}} \left( |c_x^{+}\rangle \otimes |i,j \rangle + |c_x^{-}\rangle \otimes |i +1,j\rangle \right) \,,
\label{2Dsth}
\end{eqnarray}
where the initial state  $|\psi_{in}^l \rangle = |\psi_{c}^l \rangle  \otimes |\psi_{v}\rangle$.

{\it Case 2:} Similarly, in the vertical case, the stationary  state has the following form 
\begin{eqnarray} \nonumber 
|\psi_{vstat}\rangle =   |\psi_{in}^l \rangle   \hspace{5cm}\\
-\sqrt{\frac{4+l}{N}} \left( |c_y^{+}\rangle \otimes |i,j \rangle + |c_y^{-}\rangle \otimes |i,j +1\rangle \right) \,.
\label{2Dstv}
\end{eqnarray}

{\it $U_l$:}  It has  be shown in ref.  \cite{nahi} that these two types of  states   $|\psi_{hstat}\rangle$ and $|\psi_{vstat}\rangle$ are  eigenstates of  the evolution operator  
of   lackadaisical quantum walk, $U_l = S\mathcal{C}_l$,    with unit eigenvalue:  
\begin{eqnarray}
U_l |\psi_{hstat}\rangle &=& |\psi_{hstat}\rangle\,, \\
U_l |\psi_{vstat}\rangle &=& |\psi_{vstat}\rangle \,.
\label{2Dstei}
\end{eqnarray}

{\it $U_{Grov}$:}  These two states $|\psi_{hstat}\rangle$ and $|\psi_{vstat}\rangle$  remain eigenstates  \cite{rivosh1}  for the  regular quantum walk search  with Grover coin,   
$U_{Grov}= U_{l=0}$,  
\begin{eqnarray}
U_{Grov} |\psi_{hstat}\rangle_{l=0}  &=& |\psi_{hstat}\rangle_{l=0}\,, \\
U_{Grov} |\psi_{vstat}\rangle_{l=0} &=& |\psi_{vstat}\rangle_{l=0} \,.
\label{2Dsteig}
\end{eqnarray}

{\it $U_{G}$:}  However,  under  $U_{G}= S\mathcal{C}_{G}$,    the stationary states do  not remain stationary as can be seen from the  following expressions
\begin{eqnarray} \nonumber
U_{G} |\psi_{hstat}\rangle = \hspace{6cm} \\  \nonumber
|\psi_{hstat}\rangle - \frac{2}{\sqrt{N}} S \left( |\psi_c^{x+}\rangle \otimes |i,j \rangle + |\psi_c^{x-}\rangle \otimes |i +1,j\rangle \right) \\ 
- 2\sqrt{\frac{l}{N(4+l)}} SD_l | l \rangle \otimes\left(  |i,j \rangle +  |i +1,j\rangle \right)\,, \\ \nonumber
U_{G} |\psi_{vstat}\rangle =  \hspace{6cm} \\  \nonumber
 |\psi_{vstat}\rangle  - \frac{2}{\sqrt{N}} S \left( |\psi_c^{y+}\rangle \otimes |i,j \rangle + |\psi_c^{y-}\rangle \otimes |i,j+1\rangle \right) \\
- 2\sqrt{\frac{l}{N(4+l)}} SD_l | l \rangle \otimes\left(  |i,j \rangle +  |i,j+1\rangle \right) \,.
\label{2Dstei}
\end{eqnarray}

{\it $U_{SKW}$:} Under  $U_{SKW}= S\mathcal{C}_{SKW}$ also   the stationary states do  not remain stationary as can be seen from the  following expressions
\begin{eqnarray}  \nonumber
U_{SKW} |\psi_{hstat}\rangle = \hspace{6cm} \\ 
 |\psi_{hstat}\rangle - \frac{2}{\sqrt{N}} S \left( |\psi_c^{x+}\rangle \otimes |i,j \rangle + |\psi_c^{x-}\rangle \otimes |i +1,j\rangle \right) \,, \\  \nonumber
U_{SKW} |\psi_{vstat}\rangle = \hspace{6cm} \\  
 |\psi_{vstat}\rangle - \frac{2}{\sqrt{N}} S \left( |\psi_c^{y+}\rangle \otimes |i,j \rangle + |\psi_c^{y-}\rangle \otimes |i,j+1\rangle \right) \,.
\label{2Dskw}
\end{eqnarray}
Note that, for the  quantum walk search  with SKW coin  there is no self-loop state, so we need to  set $|c_l\rangle =0$ and  $l=0$ wherever  necessary.

\subsubsection{Searching adjacent vertices} \label{exsea}
We can express the initial state $|\psi_{in}^l \rangle$  in terms of the stationary state  $|\psi_{hstat}\rangle$  as 
\begin{eqnarray} \nonumber
 |\psi_{in}^l \rangle =    |\psi_{hstat}\rangle +  \hspace{4cm}\\
 \sqrt{\frac{4+l}{N}} \left( |c_x^{+}\rangle \otimes |i,j \rangle + |c_x^{-}\rangle \otimes |i +1,j\rangle \right)\,.
 \label{inst}
\end{eqnarray}
The final state, after applying  $U_l$  repeatedly $t$ times, becomes 
\begin{eqnarray} \nonumber
U_l^t |\psi_{in}^l \rangle =    |\psi_{hstat}\rangle +  \hspace{4cm}\\
 \sqrt{\frac{4+l}{N}}  U_l^t \left( |c_x^{+}\rangle \otimes |i,j \rangle + |c_x^{-}\rangle \otimes |i +1,j\rangle \right)\,.
 \label{inst1}
\end{eqnarray}
Note that,  first part of the right hand side of  eq. (\ref{inst1}) does not change,  only the second part changes under the action of the evolution operator.  The amplitude  shifts  away from the marked vertex pair after each application of  the evolution. The upper bound of the probability of marked vertices can be  obtained  by setting $U_l^t  = -\mathbb{I}$ on the second term. 
Therefore,  success probability   of the marked vertices is bounded  as $p_{succ} \leq \mathcal{O}\left(1/N\right)$.
Note,   this  bound also holds  for  $U_{Grov}$.  The same conclusion can be drawn by taking the other stationary state   $ |\psi_{vstat}\rangle$ as well. 
It implies that  the quantum walk search by  $U_l$ and  $U_{Grov}$  on a two-dimensional grid  can not  find the adjacent vertices, because the success probability  remains bounded by the initial success probability.  The  general  configurations of the form $k \times 2m $ and   $2k \times m$, for $k,m$ being positive integers, can be obtained by  tiling   with 
basic forms $2\times 1$ and $1\times 2$,  also  can not be found by  $U_l$ and $U_{Grov}$  as well.

\begin{figure}[h!]
  \centering
     \includegraphics[width=0.40\textwidth]{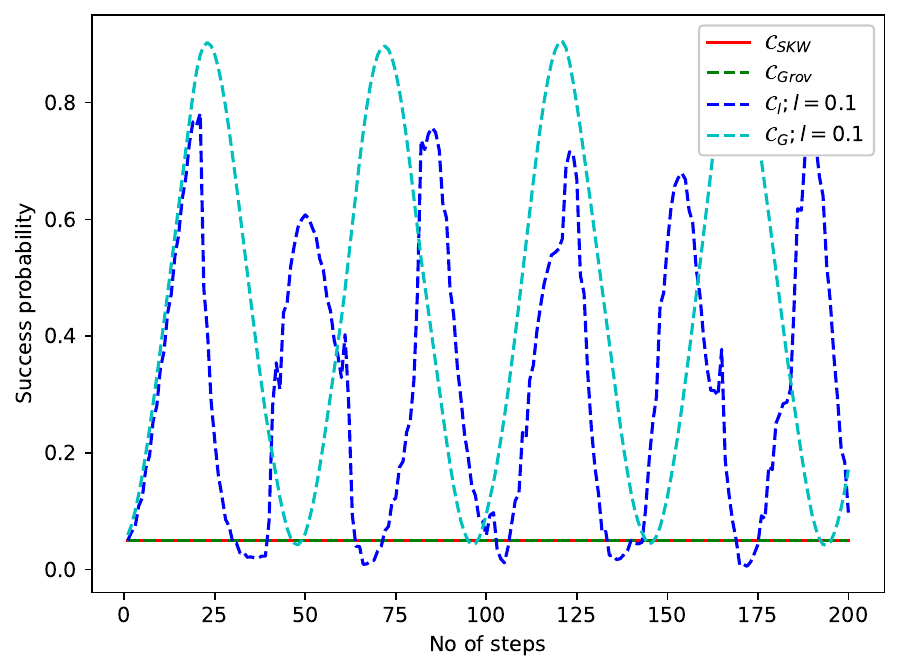}
          
       \caption{Success probability  to measure the diagonal marked vertices({\it exceptional configurations II})  on  a  $20 \times 20$ square grid as a function of the number of iteration steps   for  four different coin operators.}

\end{figure}

However,  under the action of   $U_{G}$ on  the initial state  both parts  of the initial state in eq. (\ref{inst}) transform  non-trivially, since   $ |\psi_{hstat}\rangle$ is no longer an eigenstate of  $U_{G}$.  After repeated application of the evolution operator the  final state has   significantly high and constant overlap with the marked vertices. 

For  $U_{SKW}$ also both parts of the initial state  transform  non-trivially.  However,  to obtain  high success probability we  need to  exploit amplitude amplification in addition to repeated application of $U_{SKW}$.

{\it Experimental results:}   The  behaviour of    the success probability to measure a randomly chosen pair of  adjacent marked  vertices in the horizontal direction  on a $20 \times 20$ square lattice as a function of the number of iteration steps has been presented in fig. 3.  We see that the  success probability for the $\mathcal{C}_{Grov}$ and  $\mathcal{C}_l$   coin represented by the  green and blue dashed  curves  do not grow at all even if we increase the number of iteration steps. For  the $\mathcal{C}_{G}$  and  $\mathcal{C}_{SKW}$  coin   success probability  grow  as  a function of the number of  iteration steps   represented   by the cyan and red  dashed  curves.   For the SKW coin,   as we can see,  the success probability is not significantly high, so  we need to further  apply amplitude amplification  to enhance the success probability.  The experiment is  repeated  for $10$ randomly chosen adjacent vertices in the horizontal direction and another $10$ randomly chosen adjacent vertices in the vertical  direction with  the same experimental setting, which  agrees with the result  reported in fig. 3.

We have also studied the performance of  $\mathcal{C}_{G}$   and   $\mathcal{C}_{l}$ coins   to search  the  block of marked vertices of the form $k \times m$, for $k,m$ being odd. These configurations, which  are not {\it exceptional configurations}, can be searched by the quantum walk algorithm with all the  analysed coin operators in this article. From fig. 4(top) we see that  the  success probability  to search marked vertices  by the  quantum walk algorithm with  $\mathcal{C}_{G}$  coin, given by the   blue, red and green  dashed lines  are much better compared to the   quantum walk algorithm with  $\mathcal{C}_{l}$  coin, given by the   blue, red and green  continuous  lines.   From the bottom  figure we see that  running time of  $\mathcal{C}_{G}$  can almost overlap with the standard running time given by  $\mathcal{C}_{l}$  coin  by suitably tuning the self-loop weight $l$ without much compromising  success probability as the success probability of  $\mathcal{C}_{G}$  coin changes slowing as a function of $l$. Although   $\mathcal{C}_{Grov}$  and  $\mathcal{C}_{SKW}$  coins can also  search these configurations, we have not  included these results in fig. 4, because  we need additional  amplitude amplifications technique for these two coins  to  obtain high enough success probability to compare.   

\begin{table} 
 
\begin{center}

    \begin{tabular}{ |  p{1.4cm }  | p{1.4cm } |  p{1.4cm}  | p{1.4cm}  |}
    \hline
    \multirow{2}{*}{\textbf{{\scriptsize Coin}}} & \multicolumn{2}{c|}{\textbf{ {\scriptsize Excep. configs. I}}} &   \multicolumn{1}{c|}{\textbf{ {\scriptsize Excep. configs. II }}}\\
     \cline{2-4}
     
       & \textbf{ {\scriptsize Stationary state exists?}}  & \textbf{ {\scriptsize Quantum  walk search possible?}}  &   \textbf{ {\scriptsize Quantum  walk search possible?}} \\ 
    
     \cline{2-4}
     \hline

    $\mathcal{C}_{Grov}$ & \mbox{Yes}  & \mbox{No}    &  \mbox{No}   \\ \hline
    
    $\mathcal{C}_{SKW}$  & \mbox{No}   &  \mbox{Yes}     &  \mbox{No}   \\  \hline
    
     $\mathcal{C}_l$  &   \mbox{Yes}   & \mbox{No}    &  \mbox{Yes}   \\ \hline
     \rowcolor{lightgray}    
     $ \mathcal{C}_{G}$  & \mbox{No}  & \mbox{Yes}    &  \mbox{Yes}   \\  \hline

 \end{tabular}
  \caption{Comparison of the performance of quantum walk search algorithm with four different coin operators  on a two-dimensional grid  to search for the  {\it exceptional configurations}.} 

 \end{center}
 \end{table} 

\begin{figure}[h!]
  \centering
     \includegraphics[width=0.40\textwidth]{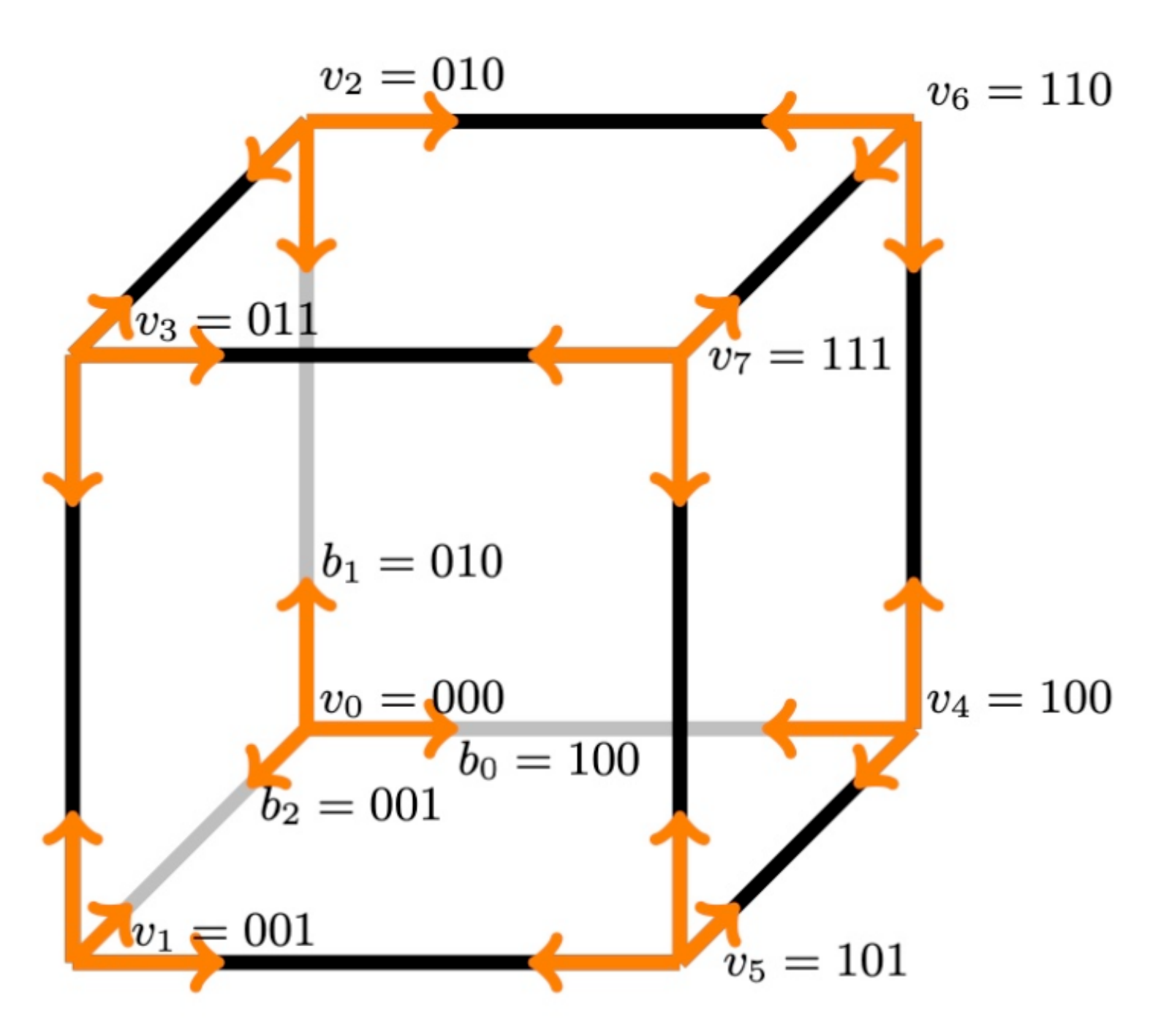}
          
       \caption{ A  $n=3$ dimensional hypercube with $N= 2^n = 8$ vertices.}
\end{figure}

\subsection{Exceptional configurations II} \label{ex1}

Another type of {\it exceptional configurations} are   the marked vertices along the diagonal of a   two-dimensional square  lattice. In fig. 2(a)   red vertices along the diagonal form this configuration, which  the SKW coin can not find \cite{amba5}. It has been  shown  in ref. \cite{men} that this diagonal configuration is   a special case of a more general {\it exceptional configurations} of the form  $i-j = \alpha$,  $\alpha = 0, 1 \cdots, \sqrt{N}-1$.  Fig.   2(a) -(d) are examples of these general  form of {\it exceptional configurations} for    $\alpha = 0, 1, 2, 3$.  Another form  of general {\it exceptional configurations} are  of the form  $i + j = \alpha$,  which can be obtained by rotating the grids in  fig.   2(a) -(d) by  $\pi/2$.  Note that, each configuration  in fig. 2 can be obtained from its left fig by translating the marked vertices by one unit to the right as depicted by the fat arrows.  Because of the periodicity  in square lattice,  translating to the  left direction also gives the same result.   

Consider a $\sqrt{N} \times \sqrt{N}$  grid with all the diagonal vertices being marked.  It has been analytically shown in refs. \cite{amba5,men} that the success probability to measure the diagonal marked vertices in case of SKW  coin  remains at its initial value despite repeated application of   $U_{SKW}$.  To summarise their  results, consider   the states of the  form 
\begin{eqnarray}  
 |\psi^+_j\rangle &=&   \frac{1}{\sqrt{2\sqrt{N}}} 
\sum_{i} \left( |c_x^{-}\rangle  + |c_y^{+}\rangle \right) \otimes |i,i+j\rangle \,,\\ 
 |\psi^-_j\rangle  &=&   \frac{1}{\sqrt{2\sqrt{N}}}  
\sum_{i} \left( |c_x^{+}\rangle  + |c_y^{-}\rangle  \right) \otimes |i,i+j\rangle\,.
\label{newst}
\end{eqnarray}
Then the  initial state  $ |\psi_{in}\rangle $ of the quantum walk can be rewritten as 
\begin{eqnarray} 
 |\psi_{in}\rangle =   \frac{1}{\sqrt{2\sqrt{N}}}  \sum_{j=0}^{\sqrt{N}-1} \left( |\psi^+_j\rangle + |\psi^-_j\rangle \right)\,.
 \label{newini}
\end{eqnarray}
{$\mathcal{C}_{SKW}$:}  For $t < \sqrt{N}$, we have 
\begin{eqnarray}  
U_{SKW}^t  \sum_{j=0}^{\sqrt{N}-1}  |\psi^+_j\rangle  &=& - \sum_{j=1}^{t}  |\psi^-_j\rangle  + \sum_{j=0}^{\sqrt{N}-t -1}  |\psi^+_j\rangle  \,, \\
U_{SKW}^t  \sum_{j=0}^{\sqrt{N}-1}  |\psi^-_j\rangle  &=& - \sum_{j=\sqrt{N}-t}^{\sqrt{N}-1}  |\psi^+_j\rangle  + \sum_{j=t+1}^{\sqrt{N}}  |\psi^-_j\rangle  \,.
\label{2Dskwe}
\end{eqnarray}
If we apply the  evolution operator  $U_{SKW}$  to the initial state   $|\psi_{in}\rangle$  for  a time period of    $x\sqrt{N} +t$     we obtain  the final state  $(-1)^x|\phi^t \rangle$, where
\begin{eqnarray} 
 |\phi^t\rangle =  - |\psi_{in}\rangle  + \sqrt{\frac{2}{\sqrt{N}}} 
 \left[  \sum_{j=t+1}^{\sqrt{N}}  |\psi^-_j\rangle    + \sum_{j=0}^{\sqrt{N}-t-1}  |\psi^+_j\rangle  \right]\,.
 \label{newfin}
\end{eqnarray}
The second part of eq. (\ref{newfin}) becomes zero for  $t = \sqrt{N}$.  The success probability to measure  the marked vertices along the diagonal   is  $1/\sqrt{N}$.  The same result on the success probability  can also be  obtained for the generalised {\it exceptional configurations}  $j-i = \alpha$ and  $i+j = \alpha$ by appropriate translation operation.

{$\mathcal{C}_{Grov}$:}  Although it has been  discussed in the literature  that the diagonal configuration  and its associated generalised {\it exceptional configurations}  can not be searched by the SKW coin, we  find   that  the same conclusion can  be drawn about the Grover coin also. 
For $t < \sqrt{N}$, we have 
\begin{eqnarray}  
U_{Grov}^t  \sum_{j=0}^{\sqrt{N}-1}  |\psi^+_j\rangle  &=&  \sum_{j=0}^{\sqrt{N}-t-1}  |\psi^+_j\rangle  - \sum_{j=\sqrt{N}-t}^{\sqrt{N}-1}  |\psi^+_j\rangle  \,, \\
U_{Grov}^t  \sum_{j=0}^{\sqrt{N}-1}  |\psi^-_j\rangle  &=& - \sum_{j=1}^{t}  |\psi^-_j\rangle  + \sum_{j=t+1}^{\sqrt{N}}  |\psi^-_j\rangle  \,.
\label{2Dgrove}
\end{eqnarray} 
If we apply the  evolution operator  $U_{Grov}$ to the initial state   $|\psi_{in}\rangle$   for a period of time  $x\sqrt{N} +t$,   we obtain  the same final state  $(-1)^x|\phi^t \rangle$, which is obtained  in SKW coin case. 

However,    in case of  $\mathcal{C}_{l}$, $\mathcal{C}_{G}$   the initial state  evolves nontrivially,  which allows us to  search  {\it exceptional configurations}  II  with constant success probability.

{\it Experimental results:}  Behaviour of the  success probability to measure the marked vertices along the diagonal, $j-i =0$, of a $20 \times 20$ square grid as a function of the number of iteration steps has been presented in fig. 5.  Generalised {\it exceptional configurations} of the forms  $j-i = \alpha$ and $i+j =\alpha$  for all values of $\alpha$ have also been numerically  checked for the same  experimental setting, which agrees with fig. 5.

We see that the success probability for the $\mathcal{C}_{SKW}$ and  $\mathcal{C}_{Grov}$   coins represented by the  red  and green dashed curves  respectively  do not grow at all even if we increase the number of iteration steps. However, for  $\mathcal{C}_{l}$  and  $\mathcal{C}_{G}$   coins  the success probability  grow  as a function of the number of  iteration  steps   represented   by the  blue and cyan  dashed  curves respectively.

A summery of  performances  for  four different coin operators  to search {\it exceptional configurations} has been presented in table 1.  $\mathcal{C}_{Grov}$ can not find both types of {\it exceptional configurations}, while  $\mathcal{C}_{SKW}$ can not find  {\it exceptional configurations II} and $\mathcal{C}_{l}$ can not find   {\it exceptional configurations I}.  However, 
$\mathcal{C}_{G}$ can find both types of {\it exceptional configurations} with constant success probability.

\begin{figure}[h!]
  \centering
     \includegraphics[width=0.40\textwidth]{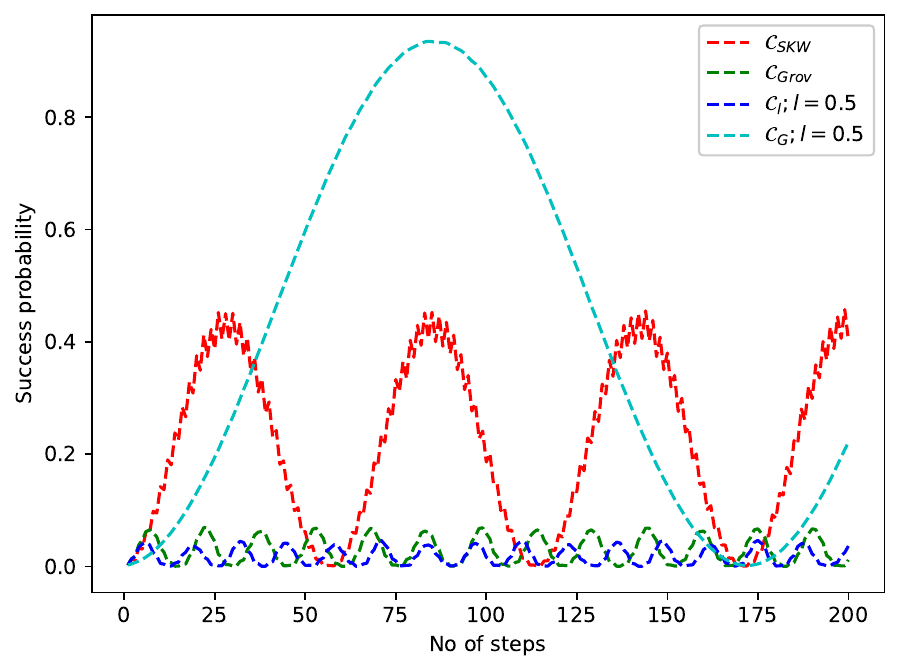}
          
       \caption{Success probability of  a randomly chosen pair of adjacent marked vertices({\it exceptional configuration I})  as a function of the number of iteration steps  for a $n=10$ dimensional  hypercube.}

\end{figure}


\section{ Quantum walk search on hypercube} \label{hyp}
Hypercube is  one of the early examples, where  discrete-time  quantum walk search has been implemented \cite{she}  to find  for a single marked vertex.  It has been shown in ref. \cite{nah1} that two adjacent  marked  vertices on the hypercube form  an {\it exceptional configuration}, which can not be found  using quantum walk search  with  Grover coin.  Here we show that,   it is possible to  search such  {\it exceptional configuration} on a hypercube by  the modified coin  $\mathcal{C}_{G}$ with high success probability. 

A  $n$-dimensional  hypercube  has  $N= 2^n$ vertices and  $n$ degree. Each vertex is connected to  $n$  nearest neighbour vertices.  We also attach one self-loop of weight $l$ to each vertex.   The Hilbert space of vertices has $2^n$ dimensions   and   the  edges  has $n+1$ dimensions.   A $3$-dimensional hypercube is depicted in fig. 6.  It has 
eight vertices and each vertex  is connected to three nearest neighbour  vertices. 
The  initial state of the coin and vertex space are 
\begin{eqnarray}
 |\psi_c^l \rangle   &=& \frac{1}{\sqrt{n+l}} \left( \sum_{x=0}^{n-1} |x \rangle + \sqrt{l}| c_l \rangle   \right),\\
 |\psi_{v}\rangle &=&   \frac{1}{\sqrt{N}}  \sum^{N-1}_{i = 0} |v_i \rangle \,,
 \label{inhy}
\end{eqnarray}
respectively, where $v_i$ is the binary representation of the decimal number $i$.  For the quantum walk search we start with the initial state $| \psi_{in} \rangle = |\psi_c^l \rangle \otimes |\psi_{v}\rangle$.  The shift operator $S$  moves the walker from a vertex to its $n$ nearest neighbour  vertices  as
\begin{eqnarray}
 S|x \rangle  |v_i \rangle   = |x \rangle  |v_i \oplus  b_x \rangle\,,
 \label{shhy}
\end{eqnarray}
where $b_x$ is the $n$-bit  binary representation of $2^x$. For example, in fig. 6, shift operation on the vertex $v_0 = 000$ becomes  $S|0 \rangle  | 000 \rangle   = |0  \rangle  |000 \oplus  100  \rangle = |0  \rangle  |100  \rangle$, $S|1 \rangle  | 000 \rangle   = |1  \rangle  |000 \oplus  010  \rangle = |1  \rangle  |010  \rangle$ and $S|2 \rangle  | 000 \rangle   = |2  \rangle  |000 \oplus  001  \rangle = |2  \rangle  |001  \rangle$.   Similarly,  the shift operator acts on all other vertices of  the hypercube as well. 

To study the exceptional configuration,  let us consider two  adjacent  marked vertices   $v_i$ and  $v_i \oplus b_x$, for a specific $i, x$.  We observe that lackadaisical quantum walk   has the following stationary  state  
\begin{eqnarray}  \nonumber 
|\psi_{stat}\rangle =   |\psi_{in}^l \rangle  \hspace{5cm}\\
-\sqrt{\frac{n+l}{N}} \left( |x \rangle \otimes | v_i \rangle + | x \rangle \otimes | v_i \oplus b_x \rangle \right).
\label{hypst}
\end{eqnarray}
After applying  $U_l$  repeatedly $t$ times, the initial state  becomes 
\begin{eqnarray} \nonumber
U_l^t |\psi_{in}^l \rangle =    |\psi_{stat}\rangle +  \hspace{4cm}\\
 \sqrt{\frac{n+l}{N}}  U_l^t \left( |x \rangle \otimes | v_i \rangle + | x \rangle \otimes | v_i \oplus b_x \rangle \right)\,.
 \label{finhyp}
\end{eqnarray}
The first part of eq. (\ref{finhyp}) does not change. Only the second part evolves under the unitary transformation. The success probability to measure the marked vertices can be found to be    $\mathcal{O}(\frac{n}{N})$ for large $n$.   Setting  $l=0$   we obtain the  result  for the Grover coin   discussed in  ref. \cite{nah1}. 

However,  it can be shown that for the $\mathcal{C}_{SKW}$ and  $\mathcal{C}_{G}$ coins  the stationary state $|\psi_{stat}\rangle$  becomes non-stationary, leading to  high success probability for the  pair of adjacent marked vertices on the hypercube. 

{\it Experimental results:}  In fig. 7,  success probability to measure  a randomly chosen  pair of adjacent  marked vertices is plotted as a function of the number of iteration steps.  We see that success probability for  $\mathcal{C}_{Grov}$ and  $\mathcal{C}_l$  represented by the  green and blue dashed curves  respectively  does  not grow  even if we increase the number of iteration steps. However, for  $\mathcal{C}_{SKW}$  and  $\mathcal{C}_{G}$  the success probability  grows   as function of the number of  iteration  steps   represented   by the red  and cyan  dashed curves  respectively.   However, for the SKW  coin  we additionally need to deploy amplitude amplification to further improve the success probability.   We observe  that   {\it exceptional configurations I}    portion  of table 1(column $2$ with two sub-columns)  agrees   with the result  of a  hypercube.    Since {\it exceptional configurations II} do not exist in a  hypercube, column $3$  of table 1 is not applicable to the hypercube. 
 The experiment is repeated for $10$ randomly chosen pairs  of adjacent  marked vertices with the same experimental setting, which agrees with  the result in fig. 7.

\section{Conclusions} \label{con}
Quantum walk  algorithm is very much successful in  searching  for a single marked vertex on various graphs.  However, searching  for multiple  marked   vertices on a graph is subject to certain limitations.  Two categories  of {\it exceptional configurations},  discussed in this article, are difficult to find by the quantum walk search  algorithm  with some of the well known coin operators.  For example, pair of  adjacent vertices  can not be found by the Grover coin based quantum walk and lackadaisical quantum walk. There exist  more general form of marked vertices in the two-dimensional grid,  $k \times 2m $ and   $2k \times m$, for $k,m$ being positive integers, which  quantum walk search   with $\mathcal{C}_{Grov}$  and $\mathcal{C}_{l}$ coins can not find.   Another type of {\it exceptional configurations}, in which  marked vertices are arranged along the diagonal of a square grid and its generalised forms, can not be found by the $\mathcal{C}_{SKW}$ and $\mathcal{C}_{Grov}$  coin based quantum walk search.  

However, these {\it exceptional configurations} can be successfully searched by  the quantum walk with $\mathcal{C}_{G}$ coin.  A  performance summary  of quantum walk search  with four different coins   presented in table 1,  and experimental results  presented in figs. 3, 5 and 7  show   that  $\mathcal{C}_{G}$ is the only coin, which can successfully search both types of  {\it exceptional configurations}.  We also analysed   {\it exceptional configurations I} on a hypercube  by numerically evaluating the success probability to measure the marked states for four coin operators,  showing   that   $\mathcal{C}_{G}$  can search marked vertices.  Note that,  although we have only considered {\it exceptional configurations} in this article, except fig. 4,    $\mathcal{C}_{G}$ coin  works for  other types of configuration  of marked vertices as well. 

We observe that, the occurrence of {\it exceptional configurations} in some of the quantum walk search algorithms  is not a quantum phenomenon. It is  merely  a limitation of  the   associated  coin operators, which can be avoided by a suitable choice of coin operator. 

We have also studied   the non-exceptional configurations of the form $k \times m$, $k,m$ being odd,  on a two-dimensional grid.
Fig. 4 shows that the performance of $\mathcal{C}_{G}$ coin  to search  these configurations  is better than the  $\mathcal{C}_{l}$ coin  over  the variation of  the self-loop weight. 

One of the limitations of our numerical  approach  is that we implemented  the action of the coin and shift operators on the quantum state in the numerical simulations, which generally requires exponential resources as pointed out in ref. \cite{por}.  
It prevents us to increase the size  of  the  graphs beyond a certain  point because of the limitations  of  the computing power.  The numerical method combined with  analytical method allowed  ref. \cite{por}  to analyse  the time complexity for multiple marked vertices(non-{\it exceptional configuration}) search on large hypercubes.  It would be interesting, if we can adopt  the analytical  approach  of refs. \cite{por, rog}  to study    {\it exceptional configurations}  with  $\mathcal{C}_{G}$  coin based quantum walk, which can improve our understanding on the searching of  {\it exceptional configurations} in the asymptotic limits.

Another   interesting future direction   could be to investigate the performance of  $\mathcal{C}_{G}$   coin  to search for marked vertices including {\it exceptional configurations} on  various other  graphs  and compare the results  with the performance of  other coin operators.

\vspace{1cm}

Data availability Statement:  The datasets generated during and/or analysed during the current study are available from the corresponding author on reasonable request.
\vspace{0.5cm}

Conflict of interest: The authors have no competing interests to declare that are relevant to the content of this article.


\end{document}